\documentclass[apjl]{emulateapj}

\def\kms{\ifmmode{\rm km\thinspace s^{-1}}\else km\thinspace s$^{-1}$\fi}
\def\ms{\ifmmode{\rm m\thinspace s^{-1}}\else m\thinspace s$^{-1}$\fi}

\shorttitle{TrES-4}

\begin{document}

\journalinfo{Accepted for publication in The Astrophysical Journal Letters}

\title{TrES-4: A Transiting Hot Jupiter of Very Low Density}

\author{Georgi Mandushev\altaffilmark{1}, Francis T. O'Donovan\altaffilmark{2},
David Charbonneau\altaffilmark{3,4}, Guillermo Torres\altaffilmark{3}, 
David W. Latham\altaffilmark{3}, G\'asp\'ar \'A. Bakos\altaffilmark{3,5}, 
Edward W. Dunham\altaffilmark{1}, Alessandro Sozzetti\altaffilmark{3,6}, 
Jos\'e M. Fern\'andez\altaffilmark{3}, Gilbert A. Esquerdo\altaffilmark{3,7}, 
Mark E. Everett\altaffilmark{7}, Timothy M. Brown\altaffilmark{8,9}, 
Markus Rabus\altaffilmark{10}, Juan A. Belmonte\altaffilmark{10}, 
Lynne A. Hillenbrand\altaffilmark{2}}

\altaffiltext{1} {Lowell Observatory, 1400 W. Mars Hill Rd., Flagstaff, AZ 
86001; gmand@lowell.edu}
\altaffiltext{2} {California Institute of Technology, 1200 E. California Blvd., 
Pasadena, CA 91125}
\altaffiltext{3} {Harvard-Smithsonian Center for Astrophysics, 60 Garden 
St., Cambridge, MA 02138}
\altaffiltext{4} {Alfred P. Sloan Research Fellow}
\altaffiltext{5} {Hubble Fellow}
\altaffiltext{6} {INAF-Osservatorio Astronomico di Torino, 10025 Pino Torinese, 
Italy}
\altaffiltext{7} {Planetary Science Institute, 1700 E. Fort Lowell Rd., 
Suite 106, Tucson, AZ 85719}
\altaffiltext{8} {Las Cumbres Observatory Global Telescope, 6740 Cortona Dr.,
Suite 102, Goleta, CA 93117}
\altaffiltext{9} {University of California, Santa Barbara, CA 93106}
\altaffiltext{10} {Instituto de Astrof{\'\i}sica de Canarias, C/ v{\'\i}a 
L\'actea s/n, 38200 La Laguna, Tenerife, Spain}

\begin{abstract}

We report the discovery of TrES-4, a hot Jupiter that transits the star
GSC~02620-00648 every 3.55~days. From high-resolution spectroscopy of the star 
we estimate a stellar effective temperature of $T_{\rm eff} = 6100 \pm 150$~K, 
and from high-precision $z$ and $B$ photometry of the transit we constrain the 
ratio of the semi-major axis $a$ and the stellar radius $R_\star$ to be 
$a/R_\star = 6.03 \pm 0.13$. We compare these values to model stellar 
isochrones to constrain the stellar mass to be 
$M_\star = 1.22 \pm 0.17$~$M_\sun$. Based on this estimate and the photometric 
time series, we constrain the stellar radius to be 
$R_\star = 1.738 \pm 0.092$~$R_\sun$ and the planet radius to be 
$R_{\rm p} = 1.674 \pm 0.094$~$R_{\rm Jup}$. We model our radial-velocity data 
assuming a circular orbit and find a planetary mass of 
$0.84 \pm 0.10$~$M_{\rm Jup}$. Our radial-velocity observations rule out 
line-bisector variations that would indicate a specious detection resulting 
from a blend of an eclipsing binary system. TrES-4 has the largest radius and 
lowest density of any of the known transiting planets. It presents a challenge 
to current models of the physical structure of hot Jupiters, and indicates that 
the diversity of physical properties amongst the members of this class of 
exoplanets has yet to be fully explored.

\end{abstract}


\section{Introduction}

Despite the ever increasing number of discovered transiting exoplanets (18 at 
the time of writing), our understanding of the relationships between host star 
properties and the planets' physical and orbital parameters is still 
incomplete. While the mass-radius relation for most transiting planets agrees 
with the models \citep{Char07a}, several planets have either too high or too 
low densities. In particular, HD209458b \citep{Knut07}, HAT-P-1b 
\citep{Bako07}, and WASP-1b \citep{Coll07} all have much larger radii and lower 
densities than expected for planets of their mass and distance from the host 
star. Several mechanisms for explaining this discrepancy have been proposed, 
among them different internal heat sources \citep{Guil02,Bode03}, increased 
planetary atmospheric opacities \citep{Burr07}, and less efficient heat 
transport as a result of interior composition gradients \citep{Chab07}. Our 
Trans-atlantic Exoplanet Survey (TrES) has previously announced the discovery 
of three transiting planets \citep{Alon04,ODon06,ODon07}, each with distinctive 
properties. We describe here the newly discovered transiting planet TrES-4, 
whose mean density of $\rho = 0.222 \pm 0.045$~${\rm g \: cm^{-3}}$ is the 
lowest of all known exoplanets with measured radii and masses. 

\section{Photometry and Spectroscopy \label{s:obs}}

We monitored a $5\fdg8 \times 5\fdg8$ field in Hercules with the Lowell 
Observatory Planet Search Survey Telescope \citep[PSST, ][]{Dunh04} and the 
Sleuth telescope at Palomar Observatory between UT 2006 May 6 and 2006 
August 2. All images were processed and the photometry and transit search 
carried out as described in \citet{Dunh04}. Both telescopes detected transits 
of the host star GSC~02620-00648: PSST observed 2 full and 1 partial transits, 
and Sleuth observed 3 full and 4 partial transits. The shape of the events 
was consistent with the transit of a Jupiter-sized planet across an F dwarf, 
and we undertook a program of followup observations to confirm the planetary 
nature of the object.

We observed the candidate with the CfA Digital Speedometer \citep{Lath92} from 
2006 September to 2007 April. We obtained 7 spectra covering 45~\AA\ centered 
at 5187~\AA, with a resolving power of $\lambda/\Delta\lambda \simeq 35\, 000$ 
and $S/N$ between 11 and 13 per resolution element. This spectral region 
includes the gravity-sensitive Mg~b triplet and the host star properties 
derived from an analysis of this region will depend on the star's metallicity. 
We obtained the radial velocities (RVs) by cross-correlation against a 
synthetic template chosen from a large library of calculated spectra based on 
Kurucz model atmospheres \citep[see][]{Nord94, Lath02}. These velocities have a 
typical precision of 0.5~\kms\ and they show no significant variation within 
the errors. We derived the effective temperature ($T_{\rm eff}$) and projected 
rotational velocity ($v \sin i$) by comparing our observed spectra against 
synthetic spectra with a wide range of parameters \cite[see][]{Torr02}. The 
values of $T_{\rm eff}$ and $v \sin i$ are listed in Table~\ref{t:host_star}, 
and assume $\left [ {\rm Fe/H} \right ] = 0.0$. We also estimate the surface 
gravity to be $\log g = 3.8 \pm 0.2$. We combined the Johnson-Cousins 
photometry that we obtained (see below) with archive 2MASS data, and used the 
color-temperature calibrations for dwarfs by \cite{Rami05} and \cite{Casa06} 
to estimate $T_{\rm eff}$. The average result, $T_{\rm eff} = 6130 \pm 80$~K, 
is consistent with the spectroscopic value. 

The mass and radius of the star were determined on the basis of the 
spectroscopic $T_{\rm eff}$, the value of $a/R_\star$ derived from the light 
curve fit described below, and an assumed metallicity of 
$\left [ {\rm Fe/H} \right ] = 0.0 \pm 0.2$. The quantity $a/R_\star$ is 
closely related to the stellar density, and is determined in this case with 
much higher relative precision than $\log g$. It is therefore a better proxy 
for luminosity \citep[see][]{Sozz07}. We searched for the best match between 
stellar evolution models interpolated to a fine grid in age and metallicity 
and the three observables, within the stated errors. Uncertainties were 
derived from the full range in $M_\star$ and $R_\star$ allowed by the 
isochrones within observational errors. A comparison with the models from the 
series by \cite{Yi01} yielded $M_\star = 1.22 \pm 0.17$~$M_\sun$, 
$R_\star = 1.738 \pm 0.092$~$R_\sun$, and $\log g = 4.045 \pm 0.034$. This 
value of $\log g$ is consistent with, but better constrained than the 
spectroscopic estimate. The age we derive is $4.7 \pm 2.0$~Gyr, and the 
predicted absolute visual magnitude ($M_V = 3.36 \pm 0.27$) implies a distance 
of $440 \pm 60$~pc, ignoring extinction. The error estimates above do not 
include possible systematic errors in the stellar evolution models. The 
gravity, radius and age estimates for the host star indicate that this is an 
evolved star at the base of the subgiant branch.

\begin{deluxetable}{llcc}
\tablewidth{0pt}
\tablecaption{TrES-4 Host Star \label{t:host_star}}
\tablehead{
\colhead{Parameter} & \colhead{Units} & \colhead{Value} & \colhead{Source}}
\startdata
RA                            & J2000.0       & $17^{\rm h} 53^{\rm m} 13\fs 05$  & 1 \\
Decl.                         & J2000.0       & $+37\arcdeg 12\arcmin 42\farcs 6$ & 1 \\
GSC                           &               & 02620-00648                       &   \\
$[\mu_{\alpha},\mu_{\delta}]$ & mas~yr$^{-1}$ & $\left [-6.5,-23.5 \right ]$      & 1 \\
$V$                           &               &     $11.592 \pm 0.004$            & 2 \\
$B-V$                         &               & \phn$ 0.520 \pm 0.007$            & 2 \\
$V-R_{\rm C}$                 &               & \phn$ 0.312 \pm 0.008$            & 2 \\
$V-I_{\rm C}$                 &               & \phn$ 0.601 \pm 0.008$            & 2 \\
$J$                           &               &     $10.583 \pm 0.018$            & 3 \\
$J-H$                         &               & \phn$ 0.233 \pm 0.024$            & 3 \\
$J-K_s$                       &               & \phn$ 0.253 \pm 0.026$            & 3 \\
$M_\star$                     & $M_\sun$      & \phn$1.22 \pm 0.17$               & 2 \\
$R_\star$\tablenotemark{a}    & $R_\sun$      & \phn$1.738 \pm 0.092$             & 2 \\
$L_\star$                     & $L_\sun$      & \phn$3.74 \pm 0.86$               & 2 \\
$T_{\rm eff}$                 & K             &     $6100 \pm 150$                & 2 \\
$\log g$               & ${\rm cm \: s^{-2}}$ & \phn$4.045 \pm 0.034$             & 2 \\
$v \sin i$                    & \kms          & \phn $9.5 \pm 1.0$                & 2 \\
Age                           & Gyr           & \phn$4.7 \pm 2.0$                 & 2 \\
Distance                      & pc            & $440 \pm 60$                      & 2 \\
\enddata
\tablenotetext{a}{This value of $R_\star$ is derived from fitting the entire
light curve and is better constrained than the value obtained solely from the
stellar evolution models (Sec.~2). The uncertainty in $R_\star$ 
includes both the statistical incertainty of $0.044 \, R_\sun$ and 
the 14\% uncertainty in $M_\star$}
\tablerefs{(1) UCAC2 \citep{Zach04}; (2) this paper; (3) 2MASS \citep{Skru06}}
\end{deluxetable}

In order to characterize the host star, we obtained off-transit 
$BV(RI)_{\rm C}$ photometry of TrES-4 on UT 2007 April 14 with the 1.05-m Hall 
telescope at Lowell Observatory in combination with a 
$2{\rm K} \times 2{\rm K}$ SITe CCD. We calibrated the photometry using 7 
standard fields from \citet{Land92}. The results are listed in 
Table~\ref{t:host_star} together with other relevant data for the host star of 
TrES-4.

\begin{figure}
\voffset=-3 cm
\epsscale{.8}
\plotone{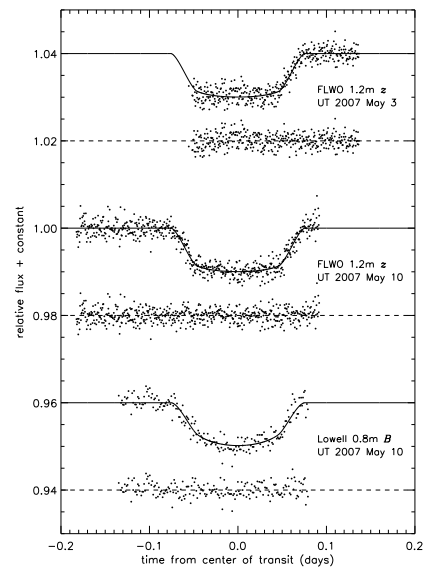}
\caption{High-precision followup $z$-band and $B$-band photometry of TrES-4. 
The plot shows the relative flux (including the color-dependent extinction 
correction) of the TrES-4 system as a function of time relative to the center 
of transit, adopting the ephemeris in Table~\ref{t:planet}. Each light curve 
is labeled with the telescope and date of observation. The residuals from the 
simultaneous fits (solid lines) are shown below each light curve.} 
\label{f:phot}
\end{figure}

We carried out high-precision in-transit $z$-band photometry of TrES-4 on 
UT 2007 May 3 and UT 2007 May 10 with KeplerCam \citep{Holm06} at the Fred L. 
Whipple Observatory (FLWO) 1.2-m telescope, and $B$-band photometry on 
UT 2007 May 10 using NASACam at the Lowell Observatory 0.8-m telescope. With 
KeplerCam we gathered 373 and 540 $z$-band 30-second exposures on UT 2007 May 3 
and UT 2007 May 10, respectively. With NASACam, we obtained 192 60-second 
$B$-band exposures. For both data sets, we derived differential fluxes relative 
to an ensemble of local comparison stars. The photometry is shown in 
Fig.~\ref{f:phot}.

We obtained high-precision RV measurements of TrES-4 on UT 2007 March 27-29, 
using HIRES and its I$_2$ absorption cell \citep{Vogt94} on the Keck~I 
telescope. Our spectra have a nominal resolving power 
$\lambda/\Delta\lambda\simeq 55\,000$ and a typical signal-to-noise ratio 
$S/N\sim 120$~pixel$^{-1}$. We used nine echelle orders in the wavelength range 
3200-8800~\AA\ to derive the velocities for TrES-4. The four star-plus-I$_2$ 
spectra (plus one I$_2$-free template exposure) provide good coverage of the 
critical phases. As the ephemeris of the system is fixed, RV data are needed 
only to determine the amplitude and systemic velocity of the orbit. In these 
cases \citep[see][]{Kona03} a few measurements are enough. We extracted and 
reduced all raw spectra using the MAKEE software written by T. Barlow 
\citep[see][]{Sozz06}. The final RV values are listed in Table~\ref{t:rv}. 
The typical precision of the 
radial velocities $\left \langle \sigma_{RV} \right \rangle \simeq 10 \, \ms$ 
is slightly degraded by the modest stellar rotation. The high-resolution, high 
$S/N$ Keck spectra were then used to rule out blend scenarios (see below), and 
they are presently being analyzed for improved characterization of the host 
star.

\begin{figure}
\epsscale{1}
\plotone{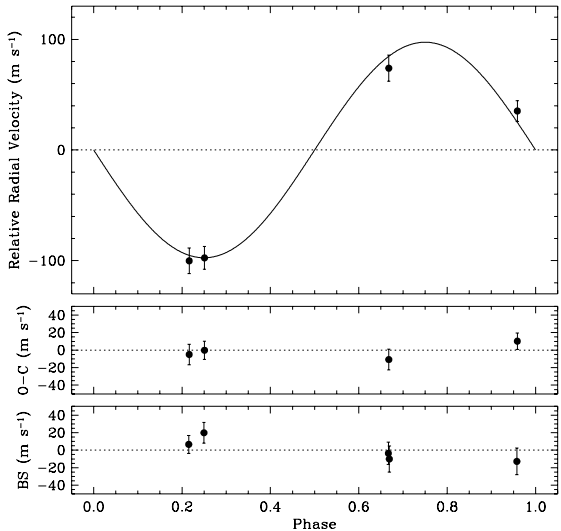}
\caption{{\em Top}: Radial velocity observations of TrES-4 obtained with 
Keck/HIRES using the I$_2$ cell, shown relative to the center of mass and 
adopting the ephemeris in Table~\ref{t:planet}. The best-fit orbit 
(solid line) is overplotted. {\em Middle}: Residuals from the best-fit 
model to the radial velocities. {\em Bottom}: Bisector spans shifted to a 
median of zero, for each of the iodine exposures as well as for the template 
(which is shown as an additional point at phase 0.669).} 
\label{f:rv}
\end{figure}

We fit a Keplerian orbit to these data assuming zero eccentricity as a good 
first approximation, as expected from theoretical arguments for a period as 
short as 3.55~days. The period and epoch of transit were held fixed. The rms of 
this fit is 11~\ms, which is similar to the internal errors of the velocities. 
The parameters of this orbital solution are listed in Table~\ref{t:planet}. We 
find 
$M_{\rm p} \sin i = 0.731 \pm 0.057~[(M_\star + M_{\rm p})/M_{\sun}]^{2/3}$~$M_{\rm Jup}$. 
The orbit is displayed in Fig.~\ref{f:rv} (top panel) along with the 
observations, and residuals are shown in the middle panel.

We investigated the possibility that the RV variations are the result of 
distortions in the line profiles caused by contamination from an unresolved 
eclipsing binary \citep{Sant02,Torr05}, instead of being caused by a planetary 
companion. We cross-correlated each Keck spectrum against a synthetic template 
matching the properties of the star, and averaged the correlation functions over 
all orders blueward of the region affected by the I$_2$ lines. From this 
representation of the average spectral line profile we computed the mean 
bisectors, and as a measure of the line asymmetry we calculated the ``bisector 
spans" as the velocity difference between points selected near the top and 
bottom of the mean bisectors \citep{Torr05}. If the RV variations were the 
result of a blend with an eclipsing binary, we would expect the line bisectors 
to vary in phase with the photometric period with an amplitude similar to that 
of the velocities \citep{Quel01,Mand05}. Instead, we detect no variation in 
excess of the measurement uncertainties (see Fig.~\ref{f:rv}, bottom panel), 
and we conclude that the RV variations are real and that the star is orbited by 
a Jovian planet.

\begin{deluxetable}{lr}
\tablewidth{0pt}
\tablecaption{Radial Velocity Measurements of TrES-4 \label{t:rv}}
\tablehead{
\colhead{HJD} & \colhead{\ \ RV (\ms)}}
\startdata
2454187.07793  &  \phm{-}$ 97.5 \pm 11.8$ \\
2454188.11122  &  \phm{-}$ 58.9 \pm \phn 9.4$ \\
2454189.02631  &  $-76.5 \pm 11.6$ \\
2454189.14920  &  $-73.9 \pm 10.3$ \\
\enddata
\end{deluxetable}

\section{Properties of TrES-4 and Discussion}

We analyzed the two $z$-band and one $B$-band photometric time series using the 
analytic light curves of \citet{Mand02}. We assumed a circular orbit and a 
quadratic stellar limb-darkening law, fixing the coefficients at the 
color-dependent values tabulated in \citet{Clar00,Clar04} for the 
spectroscopic $T_{\rm eff}$ and $\log g$, and assuming solar metallicity. We 
first estimated the time of center of transit $T_c$ by fitting 
a model light curve to the $z$ data gathered on UT 2007 May 10. We then 
determined the orbital period $P$ by combining these data with the TrES 
discovery data (which affords a baseline of 1.0 yrs). We then fixed the values 
of $T_c$ and $P$ (listed in Table~\ref{t:planet}) in the subsequent analysis.

Our model has six free parameters: the planet radius $R_{\rm p}$, the stellar 
radius $R_\star$, the orbital inclination $i$, and the color-dependent 
extinction for each of the three photometric datasets, $k_{z1}$, $k_{z2}$, and 
$k_B$. We assume that the observed flux is proportional to $e^{-km}$, where $m$ 
denotes the air mass. The values of $R_{\rm p}$ and $R_\star$ as constrained by 
the light curves alone are covariant with $M_\star$. In our analysis, we first 
estimated the quantity $a / R_\star$ which is independent of the assumed value 
of $M_\star$, and then used this estimate to constrain the value of $M_\star$ 
from stellar isochrones \citep{Sozz07,Holm07}. We then fixed 
$M_\star = 1.22$~$M_{\sun}$, and estimated the systematic error in the radii 
using the scaling relations 
$R_{\rm p} \propto R_\star \propto M_\star^{1/3}$ (see footnote to 
Table~\ref{t:host_star}).

We found first the values of $R_{\rm p}$, $R_\star$, $i$, $k_{z1}$, $k_{z2}$, 
and $k_B$ that minimized the ${\chi}^{2}$ using the AMOEBA algorithm 
\citep{Pres92}. This model is shown as the solid curves in Fig.~\ref{f:phot}. 
We then conducted a Markov Chain Monte Carlo analysis similar to that described 
in \citet{Holm06}, \citet{Char07b}, and \citet{Winn07}. We created two MCMC 
chains with 400,000 points each, one starting from the best-fit values and one 
starting from a random perturbation to those values. We then rejected the first 
23\% of the points to minimize the impact of the initial conditions and found the 
results from the two chains to be indistinguishable. We examined the histograms 
of the six parameters, as well as the histograms for several combinations of 
parameters relevant to anticipated followup studies. We assigned the optimal 
value to the median, and the 1-$\sigma$ errors to the symmetric range about the 
median than encompassed 68.3\% of the values. We list these estimates in 
Table~\ref{t:planet}. 

\begin{deluxetable}{llr@{$\: \pm \:$}l}
\tablewidth{0pt}
\tablecaption{TrES-4 Planet Parameters \label{t:planet}}
\tablehead{
\colhead{Parameter} & \colhead{Units} & \multicolumn{2}{c}{Value}}
\startdata
$P$                           & days          & $3.553945$ & $0.000075$     \\
$T_c$                         & HJD           & $2\, 454\, 230.9053$ & $0.0005$ \\
$a$                           & AU            & $0.0488$ & $0.0022$         \\
$i$                           & deg           & $82.81$  & $0.33$           \\
$a/R_\star$                   &               & $6.026$  & $0.131$          \\
$b = a \cos i/R_\star$        &               & $0.755$  & $0.015$          \\
$K$                           & \ms           & $97.4$   & $7.2$            \\
$\gamma_{_{\rm HIRES}}$       & \ms           & $+23.7$  & $5.8$            \\
$M_{\rm p}$                   & $M_{\rm Jup}$ & $0.84$   & $0.10$           \\
$R_{\rm p}$\tablenotemark{a}  & $R_{\rm Jup}$ & $1.674$  & $0.094$          \\
$\bar \rho$ & ${\rm g \: cm^{-3}}$            & $0.222$  & $0.045$          \\
$\log g$         & ${\rm cm \: s^{-2}}$       & $2.871$  & $0.038$          \\
$R_{\rm p}/R_\star$           &               & $0.09903$ & $0.00088$       \\
\enddata
\tablenotetext{a}{The uncertainty in $R_{\rm p}$ includes both the statistical 
incertainty of $0.053 \, R_{\rm Jup}$ and the 14\% uncertainty in $M_\star$ from
Table~\ref{t:host_star}}
\end{deluxetable}

TrES-4 has the largest radius and the lowest density of all exoplanets whose 
mass and radius are known, and as such presents new challenges for the theory 
of irradiated gas giants. \citet{Burr07} suggested that increased planetary 
atmospheric opacities and the inclusion of a transit radius correction 
\citep{Bara03,Burr03} can explain the radii of all large planets. It appears, 
however, that TrES-4's radius is still too large for its mass, age and 
insolation. In terms of mass and distance to the host star TrES-4 is similar 
to HD209458b, but has nearly 30\% larger radius ($R_{\rm p} = 1.67$ vs. 
$R_{\rm p} = 1.32$). Because of the higher host star luminosity TrES-4 receives 
about twice the stellar flux, but it is unlikely that the radius difference can 
be explained solely by the higher stellar irradiation. For example, HAT-P-1b, 
which has about the same radius as HD209458b and is the only other planet with 
$\rho < 0.3$~${\rm g \: cm^{-3}}$, receives only 60\% of the flux that 
HD209458b does. None of the recent models of hot Jupiters 
\citep[{\em e.g.},][]{Fort07,Burr07} can predict a radius as large as that of 
TrES-4 at its orbital separation for the estimated age and for any mass, even 
when higher atmospheric opacities are considered and allowance for the transit 
radius correction is made.

The properties of TrES-4 and its host star allow for many interesting followup 
studies, some of which are already underway. The large radius of TrES-4 in the 
visual suggests that it may have an extended outer atmosphere, similar to that 
detected around HD209458b in Lyman~$\alpha$ by \citet{Vida03}. If such an 
extended envelope is caused by mass loss through the planet's Roche lobe 
boundary, TrES-4 may have an even bigger envelope because of its smaller 
Roche limit: $R_{\rm Roche} \approx 3.5$~$R_{\rm p}$ vs. 
$R_{\rm Roche} \approx 4.4$~$R_{\rm p}$ for HD209458b \citep{Erka07}. 
Moreover, TrES-4 might be in a strong hydrodynamic ``blow-off" regime where the 
outer atmospheric layers are detached from the planet's gravitational field and 
escape in a comet-like tail \citep{Vida03,Leca04}.

The relatively fast rotation ($v \sin i = 9.5 \, \kms$) and brightness of the 
TrES-4's host star, as well as the planet's size, are favorable for measuring 
the Rossiter-McLaughlin effect (RME) \citep[{\em e.g.},][]{Gaud07}. Of 
particular interest is the angle between the star's spin axis and the planet's 
orbital plane, as it can provide information about the possible mechanisms of 
the planet migration and interaction with the protoplanetary disk. The 
semi-amplitude of the RME is proportional to $v \sin i$ and to 
$(R_{\rm p} / R_\star)^2$ and we estimate that for TrES-4 the RME can reach 
about $90~\ms$, or roughly equal to the semi-amplitude of its orbital velocity. 

\acknowledgments

We thank Travis Barman for useful discussions. This paper is based on work 
supported in part by NASA grants NNG04GN74G, NNG04LG89G, NNG05GI57G, 
NNG05GJ29G, and NNH05AB88I through the Origins of Solar Systems Program, and 
NASA Planetary Major Equipment grant N4G5-12229. We acknowledge support from 
the NASA {\em Kepler} mission under cooperative agreement NCC2-1390. Work by 
G. \'A. B. was supported by NASA through Hubble Fellowship grant 
HST-HF-01170.01-A. Observing time on Keck~I was awared by NASA.

\end{document}